\documentclass[conference]{IEEEtran}
\IEEEoverridecommandlockouts
\usepackage{cite}
\usepackage{amsmath,amssymb,amsfonts}
\usepackage{graphicx}
\usepackage{textcomp}
\usepackage{xcolor}
\usepackage{algorithm}
\usepackage[noend]{algpseudocode}

\def\BibTeX{{\rm B\kern-.05em{\sc i\kern-.025em b}\kern-.08em
    T\kern-.1667em\lower.7ex\hbox{E}\kern-.125emX}}
\begin{document}

\title{An Efficient Parallel Algorithm for finding
Bridges in a Dense Graph}

\author{\IEEEauthorblockN{1\textsuperscript{st} Ashwani Kumar}
\IEEEauthorblockA{\textit{Independent Scholar} \\
Delhi, India \\
ashwanisinghnet@gmail.com \\
0000-0002-5997-6732}
\and
\IEEEauthorblockN{2\textsuperscript{nd} Aditya Pratap Singh}
\IEEEauthorblockA{\textit{Independent Scholar} \\
Delhi, India \\
pratapaditya1997@gmail.com \\
0000-0003-4105-7241}
}

\maketitle

\begin{abstract}
This paper presents a simple and efficient approach for finding the bridges and failure points in a densely connected network mapped as a graph. The algorithm presented here is a parallel algorithm which works in a distributed environment. The main idea of our algorithm is to generate a sparse certificate for a graph and finds bridges using a simple DFS (Depth First Search). We first decompose the graph into independent and minimal subgraphs using a minimum spanning forest algorithm. To identify the bridges in the graph network, we convert these subgraphs into a single compressed graph and use a DFS approach to find bridges. The approach presented here is optimized for the use cases of dense graphs and gives the time complexity of O(E/M + Vlog(M)), for a given graph G(V,E) running on M machines.
\end{abstract}


\section{Introduction}
The bridges of a graph $G(V,E)$ are those edges which, if removed individually, will contribute to increase the connected components of graph $G$\cite{ref2}. Finding bridges in a graph has many applications in real-world systems like network bottlenecks, fault determination, vulnerabilities in a connected network and are useful for designing reliable networks. Of particular interest has been to develop algorithms for densely connected networks in real-world systems. In this direction, Alan P. Sprague and K.H. Kulkarni have presented algorithms for bridges in a parallel setting which relies heavily on the parallel prefix algorithm and is limited to interval graphs\cite{ref1}. But the algorithms for dense networks are not much worked upon, unlike sparse graphs such as C. Savage and Joseph Ja’Ja’[3], which are optimized for sparse graphs with a time complexity of $\mathcal{O}(\log{}n)$ using $\mathcal{O}(n^3)$ processors. These approaches work well for sparse graphs, but the algorithms don’t perform well for dense networks. In this paper, we design a simple and efficient algorithm that considers a simple Depth First Search (DFS) approach to be worked in a parallel setting optimized for densely connected graphs.

\subsection{Motivation}
One of the crucial facts that motivate our method to find bridges and its algorithm in sequential environment is that finding bridges in a graph is a relatively easier task. In a graph, one can use a simple depth-first search to find bridges. But in a distributed system parallelizing a DFS is not trivial and poses some challenges. In summary, it is not efficient to parallelize the idea of simple DFS to find a bridge. But we can optimize this approach for dense graphs by making several independent sparse graphs and then run our bridge finding algorithm on different machines, which is later formed into a single graph. We use this idea in our algorithm. Since in real-world scenarios of the network, graphs are dense, and this approach is usually very fast. Another advantage of such an algorithm is the simplicity of implementation in identifying the bridges in dense graphs. The idea of breaking graphs into sparse subgraphs is conducive to distributed environment.

\subsection{Our Contribution}
We design a parallel algorithm that is simple to understand and implement. The idea of making independent sparse subgraphs makes the sequential algorithm of finding bridges applicable to parallel environments. The most important aspect of our algorithm is how it is optimized for a dense network. We use certificate theorem to exploit graphs' properties and make our simple sequential algorithm efficient for a distributed system.

\subsection{Related Work}
There are several parallel algorithms for finding bridges in a graph.  For a graph $G$ with $n$ vertices and $m$ edges, Yung H. Tsin and Francis Y. Chin\cite{ref10} provided the parallel algorithm for finding bridges in a connected graph. Recent work to find articulation points that can be modified to find bridges is done by George M. Slota and Kamesh Madduri\cite{ref5} to find cut points in a large undirected sparse graph. The algorithm is based on identifying articulation points, and labeling edges using multiple connectivity queries. Alan P. Sprague and K.H. Kulkarni do another work to find bridges in a parallel setting that relies on interval graphs \cite{ref1}.

\subsection{Organisation of Paper}
The paper is organized into various sections. In Section 2, we provide the existing algorithms related to our work. In Section 3, we describe and present our approach to solving the problem and various lemmas used to get to the final idea. Section 4 describes the formal process and algorithm for our approach. The analysis of time complexity is presented in section 5. A comparative analysis is done with the existing work in section 6. We provide ideas for further improvements and optimizations in Section 7. In Section 8, concluding points are then made, which marks the end of our paper following the references section.

\section{Existing Work}
An efficient bridge algorithm is provided by Carla Savage and Joseph Ja’Ja’. It is given that for a graph $G(V, E)$ be a connected, undirected graph, an algorithm with time complexity of $\mathcal{O}(n^2\log^3{}n)$ is provided\cite{ref3}.
Yung H. Tsin and Francis Y. Chin\cite{ref10} presents a parallel algorithm to find all bridges in a connected, undirected graph in $\mathcal{O}(\frac{n}{K} + \log^2{} n)$ time with $nK(K \geq 1)$ processors. An inverted tree $S(V, E’)$ is constructed for graph $G(V, E)$ and then, with the help of (HLCA) highest lowest common ancestor bridges are computed.
Susanne E. Hambrusch \cite{ref9} paper discusses an approach to find bridges and biconnectivity on Minimum Area Meshes. The presented algorithms find the bridge-connected components in $\mathcal{O}(n^{\frac{3}{2}})$ time for a 2- dimensional mesh of $\mathcal{O}(n)$ area, both input in the form of an adjacency matrix and in the form of edges.

\section{Our Approach}
We design a simple and efficient approach to solve the problem of finding bridges in a graph by extending the DFS (Depth First Search) to find bridges in a graph. We extend the sequential algorithm for computing bridges using depth-first search (DFS) algorithm for graph $G=(V,E)$ which runs in $\mathcal{O}(V+E)$ time \cite{ref8}. The design of our algorithm is optimized for the case when there are a huge number of edges in the graph network but few nodes. Since it is a known fact that DFS is not very well suited for working in a parallel setting due to constraints with node discovery, i.e., there might be a large subtree below a node, and since one machine would be working on that node, it will be somewhat equivalent to a sequential algorithm. Although there are some ways to tackle this problem, it's still not a great idea to directly parallelize a DFS algorithm. Our algorithm works to use the multiple nodes of a cluster machine to run a DFS on a reduced graph in such a way that the result from individual machines can be later combined to get to the final result. Our approach would be similar to a divide and conquer algorithm; first, we divide our graph into sparse certificates and find a solution by distributing it on different machines, which are later combined to form the final solution.

Consider a graph $G = (V,E)$. We will find a sparse certificate $S$ where $S \subseteq E$. Also, for this sparse certificate we can say that for any set $X \subseteq V \times V $, $G(V,E \cup Y)$ = $G(V,S \cup Y)$. So we can say that a sparse certificate with an edge set S can replace the original edge set $E$ and the $2$ graphs formed respectively would be equivalent for our purpose of finding bridges\cite{ref4}.

Lemma 1. Let  there be a graph $G (V,E)$ and let the number of nodes be $n$ and number of edges be $m$. Then there exists a sparse certificate edge set $S$ for graph $G$ such that $|S| \leq 2(n-1)$ where $|S|$ is the number of edges for the sparse certificate edge set\cite{ref4}.

Now we define our approach to find the bridges in a distributed environment. We will use the paradigm of divide and conquer in which we will divide the graph $G$ edge set $E$ into $M$ random edge sets as $U_0, U_1, U_2, \cdots, U_Q$ where $E=U_0 \cup U_1 \cup U_2 \cdots U_Q$ where $M$ is the number of machines in our distributed cluster. Our algorithm will run in $Q$ phases where $Q = \log{}M+1$. We will have $M$ machines numbered as $C_0, C_1, C_2, \cdots, C_{M-1}$.

In the first phase each of the $i^{th}$ machines will be given the $i^{th}$ graph $G_i(V, U_i)$. Now on each of the machine an algorithm to find a sparse certificate will run and provide the output where $X_i$ is the sparse certificate for edge set $U_i$, also from the lemma $1$ we can say that $|X_i| \leq 2(n-1)$ where $n$ is the total number of nodes in the graph $G(V,U)$.

Now, in the second phase we will combine the solutions from the first phase where each machine gave the output with a sparse certificate graph $G_i(V, X_i)$ where $|X_i| \leq 2(n-1)$. In this phase the $C_0$ and $C_1$ machines will combine their sparse certificates $X_0$ and $X_1$ which will make the input graph as $X_0 \cup X_1$ and we can say that $|X_0 \cup X_1| \leq 4n-4$ now we will run our sparse certificate algorithm on $C_0$ machine as $C_1$ machine remains idle. The output graph of this phase would result in a sparse graph as $X’_0 = X_0 \cup X_1$ and we can also say that $|X’_0| \leq 2(n-1)$ as given in the lemma.

Therefore we can formalize a general divide and conquer approach where this algorithm would run until $Q$ phase and we will have our final sparse graph as $G(V,S)$ where $|S| \leq 2n-1$ and $S$ is the sparse certificate formed through these several phases. We can describe this process mathematically as following,

Before the $qth$ phase:

Let, $G_i = (V, U_i \cup U_{i+1} \cup U_{i+2} \cup \cdots \cup U_{i+2^q-1})$ 
        $U_{i+l} = \phi, \forall i+l \geq M$

For $q=0$, $G_i = (V, U_i)$

\begin{figure}
\includegraphics[width=0.5\textwidth]{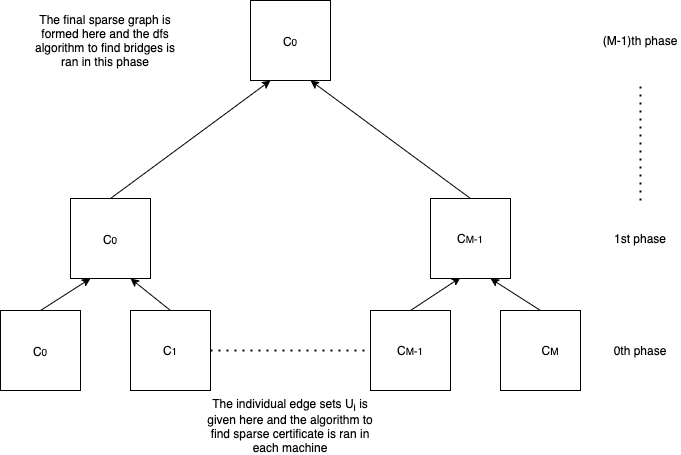} 
\caption{Phases of Algorithm}
\label{fig1}
\end{figure}

After the $i^{th}$ phase the set $X_i$ is a sparse certificate of $G_i$ and $|X_i| \leq 2(n-1)$. Also, after the $Q-1$ phases $X_0$ is the final sparse certificate of $G$ and $|X_0| \leq 2(n-1)$. Now, at this point the machine $M_0$ runs a simple DFS algorithm on the final sparse graph $G = (V,X_0)$ and computes the bridges in this graph as it follows from the statement of sparse certificates that $G(V,E \cup Y) = G(V,S \cup Y)$.
Fig \ref{fig1} depicts this process visually.

\section{Algorithm}
\subsection{General DFS algorithm}
This gives us the simple DFS algorithm for finding bridges at the final point in our sparse graph \cite{ref8}. We can see the Algorithm 1 for the formal explanation.
\begin{algorithm}
\caption{Bridge algorithm}\label{alg:euclid}
\begin{algorithmic}[1]
\Procedure{Dfs-Bridge}{$start$}{}
\State $time\gets 0$
\State $disc[start]\gets time+1$
\State $low[start]\gets time + 1$
\For{all vertex v in the graph G}
\If{there is an edge between (start, v)}
\If{v is visited}
\State $parent[v]\gets start$
\State DFS-BRIDGE(v)
\State $low[start]\gets$ min(low[start], low[v])
\If{low[v] $>$ disc[start]}
\State mark bridge from start to v
\EndIf
\Else{\textbf{If} v is not the parent of start}
\State $low[start]\gets$ min(low[start], disc[v])
\EndIf
\EndIf

\EndFor\label{euclidendwhile}
\State \textbf{done} 
\EndProcedure
\end{algorithmic}
\end{algorithm}

\subsection{Sparse certificates algorithm}
In this section we see the major algorithm to find the sparse certificates of a edge set with the help of a data structure called union-find\cite{ref6} and the subroutine DFS is used to traverse the graph, presented in Algorithm 3. We can see the Algorithm 2 for the formal explanation.

\begin{algorithm}
\caption{Sparse certificates algorithm}\label{alg:euclid}
\begin{algorithmic}[1]
\Procedure{Certificate}{}{}
\State Construct a Union-Find Data structure
\State Pick a vertex and start a DFS algorithm subroutine
\State Check for cycles using Union-Find functions and we get a graph $F$
\State $F$ is the spanning forest for graph $G = (V,E)$
\State Next find $F’$ which is the spanning forest for $G’ = (V,E-F)$
\State Let $S$ be the sparse certificate for graph $G$, where $S = F \cup F’ , S \subseteq E$
\label{euclidendwhile}
\State \textbf{done} 
\EndProcedure
\end{algorithmic}
\end{algorithm}

\begin{algorithm}
\caption{Subroutine - DFS algorithm}\label{alg:euclid}
\begin{algorithmic}[1]
\Procedure{Dfs}{$G,V$}{}\Comment{v is the source vertex}
\State $Stack S\gets \{\}$\Comment{start with an empty stack}
\State $push S \gets $v
\State DFS-BRIDGE(v)
\While{S is not empty}
\State $u\gets pop S$
\If{not visited[u]}
\State $visited[u] \gets true$
\For{each unvisited neighbour w of u}
\State $push S \gets w$
\EndFor
\EndIf
\EndWhile\label{euclidendwhile}
\State \textbf{done} 
\EndProcedure
\end{algorithmic}
\end{algorithm}

\section{Runtime Analysis}
DFS algorithm for finding bridges in an undirected graph $G(V,E)$ is a sequential algorithm which takes $\mathcal{O}(V+E)$ time on a single machine\cite{ref8}.
The algorithm to find Sparse certificates in a graph $G = (V,E)$ uses a data structure called disjoint union set which provides function to detect cycles in $\mathcal{O}(V)$ and the final algorithm constructed above runs in a $\mathcal{O}(V+E)$ time complexity on a single machine.

Further, we will discuss the time complexity for our algorithm in a distributed system environment.

Our algorithm to find sparse certificates takes $\mathcal{O}(V+E)$ time where in our approach we found a method to make the reduced set $E$ as $S$ where $|S| \leq 2(|V|-1)$ so we can say that the time complexity is $\mathcal{O}(V)$ for any $q^{th}$ phase, where $q \neq 0$.
For $q=0$ this would be given as $\mathcal{O}(V+\frac{E}{M})$ where $M$ is the number of machines.
Consequently, the first phase takes $\mathcal{O}(V + \frac{E}{M})$ time on each machine and each of the $Q-1$ remaining phases runs in $\mathcal{O}(V)$ time on each machine. Thus, the time complexity of this solution is $\mathcal{O}( V + \frac{E}{M} + V(Q-1) ) = \mathcal{O}(VQ + \frac{E}{M})$ where $Q = \log{}(M+1)$ therefore the final time complexity of the solution is $\mathcal{O}(\frac{E}{M} + V\log{}(M))$. 

\begin{figure}
\includegraphics[width=0.5\textwidth]{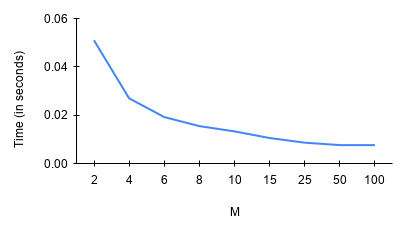} 
\caption{M is number of machines, $\lvert E \rvert$ = 10000000, $\lvert V \rvert$ = 100000}
\label{fig2}
\end{figure}

\begin{figure}
\includegraphics[width=0.5\textwidth]{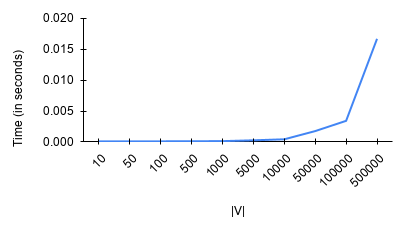} 
\caption{$\lvert V \rvert$ is number of vertex, $\lvert E \rvert$ = 100000, M = 10}
\label{fig3}
\end{figure}

\begin{figure}
\includegraphics[width=0.5\textwidth]{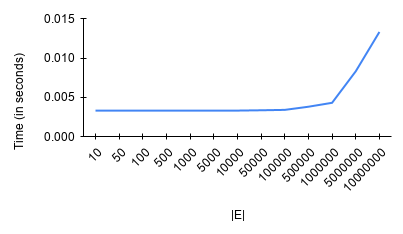} 
\caption{$\lvert E \rvert$ is number of edges, $\lvert V \rvert$ = 100000, M = 10}
\label{fig4}
\end{figure}

Fig \ref{fig2}, Fig \ref{fig3} , Fig \ref{fig4} depicts how our algorithm works for different variables in the system. It shows the behaviour of the algorithm presented here when each of the parameters are changed.  

\section{Comparative Analysis}
\begin{figure}
\includegraphics[width=0.5\textwidth]{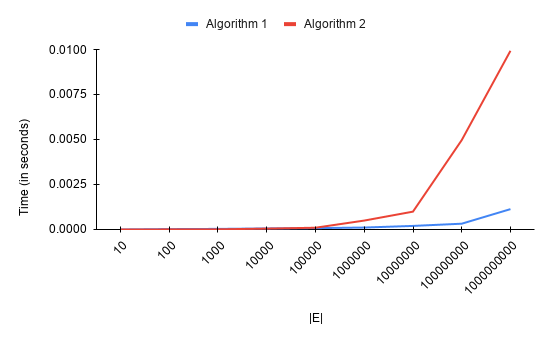} 
\caption{Our Algorithm V/S Previous Work}
\label{fig5}
\end{figure}

For the comparative analysis we picked the algorithm for finding bridges by Carla Savage and Joseph Ja’Ja’\cite{ref3} and we can see from the Fig \ref{fig5} that the algorithm presented here works really well for dense graphs and eclipses the other algorithm as we go on increasing the number of edges in the graph.

\section{Further Improvements}
For further improving the solution provided in this paper we can go on to parallelize it further. The algorithm that we are using to find the sparse certificate is currently a sequential approach. This algorithm can be implemented in a parallel setting to further optimize our approach. There is some notable work done to improve union find algorithms in a distributed environment such as given by Fredrik Manne and Md. Mostofa Ali Patwary\cite{ref7}.

\section{Conclusions}
In the paper, a unique parallel approach for identifying the bridges in a densely connected graph for a distributed environment is presented. The runtime analysis shows promising result for densely connected networks and graphs when compared with other works done in this field.

\end{document}